\documentclass{aa}

\usepackage{graphicx}
\usepackage{txfonts}
\usepackage{empheq}
\usepackage{amsmath}    
\usepackage{amssymb}    
\usepackage{bm}
\usepackage{xcolor}
\usepackage{nccmath}

\newcommand{\ev}[1]{\left\langle #1 \right\rangle}
\newcommand{\dd}{\mathrm{d}}

\begin{document} 

   \title{Fast analytical calculation of the random pair counts for realistic survey geometry}
   \titlerunning{Fast $RR$ and $DR$ calculation}

   \author{Michel-Andr\`es Breton \inst{1} \and Sylvain de la Torre \inst{1}} 

   \institute{\inst{1}Aix Marseille Univ, CNRS, CNES, LAM, Marseille, France \\
              \email{michel-andres.breton@lam.fr}
   }      


 
  \abstract
  {Galaxy clustering is a standard cosmological probe that is commonly analysed through two-point statistics. In observations, the estimation of the two-point correlation function crucially relies on counting pairs in a random catalogue. The latter contains a large number of randomly distributed points, which accounts for the survey window function. Random pair counts can also be advantageously used for modelling the window function in the observed power spectrum. Since pair counting scales as $\mathcal{O}(N^2)$, where $N$ is the number of points, the computational time to measure random pair counts can be very expensive for large surveys. In this work, we present an alternative approach for estimating those counts that does not rely on the use of a random catalogue. We derived an analytical expression for the anisotropic random-random pair counts that accounts for the galaxy radial distance distribution, survey geometry, and possible galaxy weights. We show that a prerequisite is the estimation of the two-point correlation function of the angular selection function, which can be obtained efficiently using pixelated angular maps. 
  Considering the cases of the VIPERS and SDSS-BOSS redshift surveys, we find that the analytical calculation is in excellent agreement with the pair counts obtained from random catalogues. The main advantage of this approach is that the primary calculation only takes a few minutes on a single CPU and it does not depend on the number of random points. Furthermore, it allows for an accuracy on the monopole equivalent to what we would otherwise obtain when using a random catalogue with about 1500 times more points than in the data at hand. We also describe and test an approximate expression for data-random pair counts that is less accurate than for random-random counts, but still provides subpercent accuracy on the monopole. The presented formalism should be very useful in accounting for the window function in next-generation surveys, which will necessitate accurate two-point window function estimates over huge observed cosmological volumes.}


   \maketitle
%

\section{Introduction}

The spatial distribution of galaxies has a long history of providing cosmological parameter constraints \citep[e.g.][and references therein]{strauss1992redshift, vogeley1992large, maddox1996apm,peacock01, cole05,tegmark06,percival10, blake12,delatorre2013vipers,alam2017clustering, alam2020completed}. This arises from the fact that the statistical properties of galaxies, particularly spatial ones, can be predicted by cosmological models. When analysing galaxy clustering, we usually compress the information by using summary statistics, the most natural one being the two-point correlation function or its Fourier counterpart in the power spectrum. This is due to the nearly Gaussian nature of primordial matter perturbations, which are almost fully described by their two-point statistics. Although gravitational evolution leads to non-Gaussianity, and in turn, non-vanishing higher-order $n$-point statistics, two-point statistics continues to be very informative. 

Despite the cosmological principle that implies that the correlation function is isotropic, meaning that it is only a function of the norm of the separation vector, because of the way the line-of-sight distance is measured in redshift surveys and the presence of peculiar velocities, the observed correlation function becomes anisotropic. These velocities are induced on large scales by the coherent convergence of matter towards overdensities as part of the general process of structure growth. This anisotropy makes observed galaxy $n$-point statistics sensitive to the strength of gravity acting on the large-scale structure \citep{kaiser1987clustering,guzzo2008test}.

Formally, the two-point correlation function is the excess probability of finding a pair of objects at a given distance, with respect to the expectation in a random Poisson distribution of points. In practice, we rely on statistical estimators to measure the correlation function from galaxy survey data. The first estimator was proposed by \cite{peebles1974statistical}, taking the form $\xi_{\rm PH}(\bm{s}) = DD(\bm{s})/RR(\bm{s}) - 1$, where $DD$ and $RR$ are the normalised number of distinct pairs separated by a vector $\bm{s}$, in the data and random samples, respectively. The latter sample is constructed such that random points follow the same radial and angular selection functions as the data. Other estimators were later proposed \citep{hewett1982correlation, davis1983survey, hamilton1993towards} to reduce the estimation variance, notably induced by discreteness and boundary effects. In particular, the \cite{landy1993bias} minimum-variance estimator was designed such that for any survey geometry, its variance is nearly Poissonian. This estimator, defined as
\begin{equation}
\xi_{\rm LS}(\bm{s}) = \frac{DD(\bm{s})-2DR(\bm{s})+RR(\bm{s})}{RR(\bm{s})},
\end{equation}
makes use of additional data-random pairs $DR$. To estimate the correlation function, we therefore need to compute the number of pairs as a function of the separation. To avoid introducing bias in the estimator and to minimise variance, the random catalogue must be much larger than the data catalogue \citep{landy1993bias,keihanen2019estimating}. We usually consider that taking at least about 20-50 times more random points than objects in the data is enough to avoid introducing additional variance \citep[e.g.][]{samushia2012,delatorre2013vipers,sanchez2017,bautista2021}. A problem is that the computational time for direct pair counting scales as $\mathcal{O}(N^2)$, with $N$ the number of elements in a given sample. Nonetheless, the complexity can be reduced, at best, to $\mathcal{O}(N)$ using appropriate algorithms and various efficient codes that implement them have been developed \citep[e.g.][]{moore2001fast,jarvis2004skewness, alonso2012cute, hearin2017forward, marulli2016cosmobolognalib, slepian2016accelerating, sinha2020corrfunc}. For the random-random pairs calculation specifically, additional strategies can be used to speed up the computation beyond parallelisation, such as splitting the random sample and averaging the counts over subsamples \citep{keihanen2019estimating}. Still, for large surveys, the computational time for estimating the correlation function, especially random-random pairs, can become an issue. In particular, next-generation surveys such as Euclid \citep{laureijs2011euclid} or DESI \citep{desi2016desi}, will necessitate random samples as large as about $3\times10^8$ objects in several redshift bins, or even larger if we consider a single, wide-redshift bin \citep[e.g.][]{mueller2019optimizing, castorina2019redshift}.

The role of random-random pair counts in the correlation function estimator is to account for the survey selection function, that is, the effective observed volume and its impact on the data-data pair counts. In Fourier space instead, common estimators for the power spectrum \citep[e.g.][]{feldman1994power,yamamoto2006measurement} provide a direct estimate of the window-convolved power spectrum. To be able to compare theoretical predictions to observations, it is necessary to convolve the model power spectrum with the survey window function. This convolution is computationally expensive in the likelihood analysis, but it can be done efficiently by performing a multiplication in configuration space, as proposed by \cite{wilson2017rapid}. The latter shows that the window-convolved power spectrum multipoles moments $\hat{P_\ell}(k)$ can be written as:
\begin{equation} \label{eq:windowedcf}
    \hat{P_\ell}(k) = \mathcal{H}\left[\sum_{p,q} A^q_{\ell p} \frac{2\ell+1}{2q+1} \xi_\ell(s) \mathcal{N} \frac{RR_\ell(s)}{2\pi s^2 \Delta s}\right],
\end{equation} 
where $\mathcal{H}$ denotes the Hankel transform, $A^q_{\ell p}$ are coefficients, $\xi_\ell(s)$ are model correlation function multipole moments, $\mathcal{N}$ is a normalisation factor, $RR_\ell(s)$ are the multipole moments of the random-random pair counts, and $\Delta s$ is the bin size in $s$ \citep{wilson2017rapid,beutler2017clustering}.

Random-random pairs counts are a purely geometrical quantity that depends on cosmology only through the radial selection function, which is defined in terms of the radial comoving distance. 
In the case of a simple geometry, such as a cubical volume with constant number density and periodic boundary conditions, the $RR$ pair counts can be predicted from the appropriately normalised ratio between the spherical shell volume at $s$ and the total volume. In the case of a realistic survey geometry, and taking advantage of the fact that radial and angular selection functions are usually assumed to be uncorrelated, \citet{demina2018computationally} developed a semi-analytical method to compute the $RR$ and $DR$ pair counts along the directions parallel and transverse to the line of sight, but still using a random sample to account for angular correlations. In fact, the process of spectroscopic observation can break this assumption, making radial and angular selection functions partially correlated. Nonetheless, in practice, as in the case of fibre or slit collision, for instance, we can assume independence in building $RR$ but while applying an object or pairwise weighting scheme to account for those correlations \citep[e.g.][]{delatorre2013vipers,bianchi2017,ross2017}.

In this paper, we provide general expressions for the anisotropic $RR$ and $DR$ pair counts in the case of a realistic survey geometry, including the cases for the different definitions of the pair's line of sight. We apply this formalism to the VIMOS Public Extragalactic Redshift Survey \citep[VIPERS,][]{guzzo14vimos,garilli14} and Sloan Digital Sky Survey Baryon Oscillation Spectroscopic Survey \citep[SDSS-BOSS,][]{eisenstein11,dawson13baryon} and we perform an assessment of the accuracy of the method.

This paper is organised as follows. Section 2 presents the formalism for random-random and data-random pair counts. This formalism is applied and its accuracy assessed in Section 3. We conclude in Section 4.    

\section{Formalism}

In this section, we provide the analytical formalism for the random-random and data-random pair counts.

\subsection{Random-random pairs}
\label{sec:RR_pairs}
In a survey sample where sources are selected in redshift, 
the number of sources in a given radial distance interval $[r_{\rm min}, r_{\rm max}]$ is:
\begin{equation}
    N(r_{\rm min}, r_{\rm max}) = \int_{r_{\rm min}}^{r_{\rm max}}p(r)\dd r,
    \label{eq:Nr}
\end{equation}
with $p(r)$ the number of sources as function of the radial distance $r$ and
\begin{equation}
   p(r) = r^2 n(r)\int_0^\pi \sin(\theta)\int_0^{2\pi}W(\theta, \varphi)\dd\theta\dd\varphi,
\end{equation}
where $W(\theta, \varphi)$ is the survey angular selection function in spherical coordinates. The latter encodes the probability of observing a source at any angular position on the sky and takes values from 0 to 1. Here, $n(r)$ is the source number density given by
\begin{equation}
n(r) = 
\left\lbrace
\begin{array}{lcl}
0  &  \rm{for} & r < r_{\rm min},  \\
\frac{p(r)}{4\pi r^2\ev{W}} & \rm{for} & r_{\rm min} < r < r_{\rm max}, \\
0  & \rm{for} & r > r_{\rm max},
\end{array}\right.
\label{eq:ntilde}
\end{equation}
with $\ev{W}$ the angular selection function averaged over the full sky. We note that radial weights, such as those of \citealt{feldman1994power}, can be included straightforwardly in the $p(r)$, such that this becomes a weighted radial distribution in the equations. The total number of observed sources is therefore: 
\begin{multline}
N(r_{\rm min}, r_{\rm max}) = \int_{r_{\rm min}}^{r_{\rm max}}r^2 n(r)\int_{0}^{\pi}\sin(\theta)\int_{0}^{2\pi}W(\theta, \varphi)\dd\varphi\dd\theta\dd r .
\end{multline}
In $RR(\bm{s})$, we correlate points at two different positions $\bm{r}_1$ and $\bm{r}_2$ and it is convenient to write
\begin{eqnarray}
  r_2(r_1, s, \mu) = r_1\sqrt{1+2\mu \frac{s}{r_1} + \left(\frac{s}{r_1}\right)^2},    
\end{eqnarray}
with $\bm{s} = \bm{r}_2 - \bm{r}_1$ and $\mu = \bm{r_1}\cdot\bm{s}/r_1s$. $RR$ is obtained by integrating the angular and radial selection functions first over ($r_1, \theta, \varphi$) and then over the volume defined by the separation ($s, \tilde{\theta}, \tilde{\varphi}$) as:
\begin{multline}
 RR(s_{\rm min}, s_{\rm max}) = 
  \int_{r_{\rm min}}^{r_{\rm max}}r_1^2 n(r_1)\int_{s_{\rm min}}^{s_{\rm max}}s^2\int_{0}^{\pi}\sin{\theta}\int_{0}^{\pi}\sin{\tilde{\theta}} \\
  \int_{0}^{2\pi}\int_{0}^{2\pi} n(r_2)W(\theta_1, \varphi_1)W(\theta_2, \varphi_2)\dd r_1\dd s\dd\theta\dd\tilde{\theta}\dd\varphi\dd\tilde{\varphi},
\end{multline}
where $\theta_1, \varphi_1$ ($\theta_2$, $\varphi_2$) are the angular positions at $\bm{r_1}$ ($\bm{r_2}$). 
Let us define $\hat{\bm{r}}_1 = \bm{r_1}/r_1$ and $\hat{\bm{r}}_2 = \bm{r_2}/r_2$, we have then
\begin{equation}
\int_{4\pi}\dd^2\hat{\bm{r}}_1W(\hat{\bm{r}}_1)W(\hat{\bm{r}}_2) = \int_{4\pi}\dd^2\hat{\bm{r}}_1W(\hat{\bm{r}}_1 )W(\hat{\bm{r}}_1+\bm{\phi}),
\end{equation}
and the correlation function of the angular selection function is
\begin{equation}
    \omega(\bm{\phi}) = \ev{W(\hat{\bm{r}}_1)W(\hat{\bm{r}}_1+\bm{\phi})} = \frac{1}{4\pi}\int_{4\pi}\dd^2\hat{\bm{r}}_1W(\hat{\bm{r}}_1)W(\hat{\bm{r}}_1+\bm{\phi}).
\end{equation}
In our case, since we auto-correlate randoms points, $RR$ will only depend on the angular separation (for cross-correlations with different angular selection functions, we would need to keep the angular dependence). This means that we can write $\omega(\bm{\phi}) = \omega(\phi)$, where we have
\begin{equation}
    \phi(r_1, r_2, s) = \arccos\left[\frac{r_1^2+r_2^2-s^2}{2r_1r_2}\right].
\end{equation}
We note that in the absence of an angular mask, that is, when we evenly probe the full sky, $\omega(\phi) = 1$. In compiling these results, we find that:
\begin{multline}
  RR(s_{\rm min}, s_{\rm max}, \mu_{\rm min}, \mu_{\rm max})  = \\
  8\pi^2\int_{r_{\rm min}}^{r_{\rm max}}r_1^2 n(r_1)\int_{s_{\rm min}}^{s_{\rm max}}s^2\int_{\mu^*_{\rm min}}^{\mu^*_{\rm max}} n(r_2)\omega(\phi)\dd r_1\dd s\dd\mu.
  \label{eq:final_RR}
\end{multline}
We note that we have implicitly assumed an `end-point' definition for the pair line of sight, that is, for every separation, the direction of the line of sight coincides with that of $\bm{r_1}$. With this definition, we can just use $(\mu^*_{\rm min},\mu^*_{\rm max}) = ( \mu_{\rm min},\mu_{\rm max})$ for the integral limits in Eq.~\eqref{eq:final_RR}. In the case of the `mid-point' definition for the pair line of sight, where $\mu = \bm{r}\cdot\bm{s}/rs$ with $\bm{r} = \frac{1}{2}(\bm{r_1}+\bm{r_2})$, we can use the same equation, but we need to change the integral limits for each \{$r_1, s, \mu$\} as:
\begin{equation}
    \mu^*(r_1, s, \mu) = \frac{-s+s\mu^2+\mu\sqrt{s^2\mu^2-s^2+4r_1^2}}{2r_1}.
\end{equation}
We note that for the end-point definition it is important to compute pairs with $\mu < 0$ since the correlation function in that case is not symmetric by pair exchange. For applications involving random-random multipole moments directly, the latter can be defined as: 
\begin{multline}
  RR_\ell(s_{\rm min}, s_{\rm max})  = \\
  8\pi^2\frac{(2\ell+1)}{2}\int_{r_{\rm min}}^{r_{\rm max}}r_1^2 n(r_1) \int_{s_{\rm min}}^{s_{\rm max}}s^2 \int_{-1}^{1} n(r_2)\omega(\phi)\mathcal{L}_\ell(\mu^\dagger)\dd r_1\dd s\dd\mu,
  \label{eq:final_RR_multipoles}
\end{multline}
with $\mathcal{L}_\ell$ the Legendre polynomial of order $\ell$. For the end-point definition, $\mu^\dagger = \mu$, while for the mid-point definition we have
\begin{equation}
    \mu^\dagger(r_1, s, \mu) = \mu \frac{r_1}{r} + \frac{1}{2}\frac{s}{r},
\end{equation}
with $r(r_1, s, \mu) = r_1\sqrt{1+\mu s/r + s^2/(2r)^2}$.
Finally, we note that in order to cross-correlate tracers with different radial selection functions but the same angular selection function, we can use the same formalism but with different $n_{\rm A}(r_1)$ and $n_{\rm B}(r_2)$ in Eqs.~\eqref{eq:final_RR}-\eqref{eq:final_RR_multipoles}.

\subsection{Data-random pairs}
\label{sec:DR_pairs}

The \citet{landy1993bias} estimator includes data-random pairs to minimise variance. A similar formalism to that used for random-random pair counts can be employed to evaluate data-random pair counts. However, contrarily to the $RR$ case, we now have to cross-correlate a discrete set of sources with a continuous random distribution. The discrete limit of Eq.~\eqref{eq:Nr} is
\begin{equation}
    N(r_{\rm min}, r_{\rm max}) = \sum_{i=0}^N \int_{r_{\rm min}}^{r_{\rm max}}\int_{4\pi}\delta_{\rm D}(\bm{r}-\bm{r}_i)\dd r\dd^2\hat{\bm{r}},
    \label{eq:Nr_discrete}
\end{equation}
with $\delta_{\rm D}$ the Dirac delta function, $\bm{r} = (r, \theta, \varphi)$, and $\bm{r}_i$ the $i^{\rm th}$ source position in the data vector. We can then use the same methodology as in Section~\ref{sec:RR_pairs}. To make the computation of the data-random pair counts tractable we make the assumption:
\begin{multline}
     \sum_{i=0}^N\int_{r_{\rm min}}^{r_{\rm max}} \dd r_1\int_{4\pi} \dd^2\hat{\bm{r}}_1\delta_{\rm D}(\bm{r}_1-\bm{r}_i)W(\hat{\bm{r}}_1+\bm{\phi})  \approx \frac{1}{N} \sum_{i=0}^N \sum_{j=0}^N \\
   \int_{r_{\rm min}}^{r_{\rm max}} \dd r_1\delta_{\rm D}(r_1 - r_i) \int_{4\pi} \dd^2\hat{\bm{r}}_1\delta_{\rm D}(\hat{\bm{r}}_1-\hat{\bm{r}}_j)W(\hat{\bm{r}}_1+\bm{\phi}), 
  \label{eq:approx_DR}
\end{multline}
with $\hat{\bm{r}}_1 = (\theta, \varphi)$ and $\hat{\bm{r}}_i$ the angular position in the data vector. Under this assumption, the angular correlation at a particular point is given by that of the whole sample. For a large-enough $N$ and a sufficiently homogeneous angular sampling of the data, the approximation should hold. We find that:
\begin{multline}
    DR(s_{\rm min}, s_{\rm max}, \mu_{\rm min}, \mu_{\rm max}) =   \sum_{i=0}^N \\
    \frac{4\pi}{N} \int_{r_{\rm min}}^{r_{\rm max}}\delta_{\rm D}(r_1-r_i)\int_{s_{\rm min}}^{s_{\rm max}}s^2 
    \int_{\mu^*_{\rm min}}^{\mu^*_{\rm max}}n(r_2)\int_0^{2\pi}\omega_{DR}(\bm{\phi})\dd r_1\dd s\dd\mu\dd\varphi,
\end{multline}
with
\begin{equation}
     \omega_{DR}(\bm{\phi}) =  \frac{1}{4\pi} \sum_{j=0}^N\int_{4\pi}\dd^2\hat{\bm{r}}_1\delta_{\rm D}(\hat{\bm{r}}_1-\hat{\bm{r}}_j)W(\hat{\bm{r}}_1+\bm{\phi}),
\end{equation}
%
where we integrate over all the data and angular selection function positions.
In practice, we can cross-correlate a map containing the galaxy number density per pixel with the angular selection function map. Introducing a generic data weight ,$w_i$, for each source and assuming that the angular cross-correlation function does not depend on the pair orientation, we obtain
\begin{multline}
  DR(s_{\rm min}, s_{\rm max}, \mu_{\rm min}, \mu_{\rm max})  =   \frac{8\pi^2}{N}\sum_{i=0}^N \\ \int_{r_{\rm min}}^{r_{\rm max}}\delta_{\rm D}(r_1-r_i)  w_i\int_{s_{\rm min}}^{s_{\rm max}}s^2\int_{\mu^*_{\rm min}}^{\mu^*_{\rm max}} n(r_2)\omega_{\rm DR}(\phi) \dd r_1\dd s\dd\mu.
  \label{eq:DR_final}
\end{multline}
We can already see that the calculation involves a sum over all data sources, which is potentially more computationally expensive to evaluate than in the $RR$ and direct pair-counting cases. Nonetheless, to make the computation efficient, we can approximate this by taking the continuous limit on the sum and write it out similarly as we do for the $RR$ case:
\begin{multline}
  DR(s_{\rm min}, s_{\rm max}, \mu_{\rm min}, \mu_{\rm max})  = \\
  8\pi^2 \int_{r_{\rm min}}^{r_{\rm max}}r_1^2 n_1(r_1)\int_{s_{\rm min}}^{s_{\rm max}}s^2\int_{\mu^*_{\rm min}}^{\mu^*_{\rm max}}n_2(r_2)\omega_{\rm DR}(\phi)\dd r_1\dd s\dd\mu,
  \label{eq:final_DR_approx}
\end{multline}
where the angular cross-correlation function, $\omega_{\rm DR}$, can be written as $\omega_{\rm DR} = \ev{W_1W_2}$, with $W_1$ as a pixelated map containing (weighted) source counts and $W_2$ as the angular selection function as in Section~\ref{sec:RR_pairs}. In this case, the normalisation on $W_1$ and $W_2$ does not matter since the final result is proportional to $\frac{\ev{W_1W_2}}{\ev{W_1}\ev{W_2}}$. In principle, Eq.~\eqref{eq:final_DR_approx} should be close to Eq.~\eqref{eq:DR_final} when the $p(r)$ is fine enough to faithfully reproduce data radial overdensities.

\section{Application} \label{sec:application}

In this section, we apply our formalism for $RR$ and $DR$ pair counts to the case of BOSS and VIPERS redshift surveys and test its accuracy.

\subsection{Numerical implementation} \label{sec:numerical_implementation}

The method takes as input the radial distance distribution of sources and the correlation function of the angular selection function. To compute the latter, we first produce a \textsc{Healpix} \citep{gorski2005healpix} full-sky map from the survey angular mask, which is generally in form of a list of distinct spherical polygons with associated weights. We infer the angular correlation function from the maps using \textsc{Polspice} \citep{szapudi2001fast, chon2004fast}, which takes advantage of {\sc Healpix} isolatitude pixelation scheme to quickly evaluate the angular correlation function $\omega(\theta)$ with fast spherical harmonic transforms. The CPU time to compute the angular correlation function depends on the map resolution, but it is generally quite efficient. The map resolution is controlled by the $nside$\footnote{The total number of pixels in a full-sky map is given by $N_{\rm pix} = 12\times \rm{nside}^2$.} parameter. For instance, for BOSS it takes 3, 20 and 130 minutes on a single CPU for $nside$ = 2048, 4096, and 8192, respectively. For VIPERS, it takes 6, 40, and 320 minutes for $nside$ = 8192, 16384, and 32768, respectively. We note that we compute the angular power spectra until $\ell_{\rm max} = 3\times nside$;  it is possible to reduce the run time by reducing $\ell_{\rm max}$. The main limitation is memory since this operation needs to load the full-sky map, which can be difficult for high-resolution maps. Once the correlation function of the angular mask, average map value $\ev{W}$, and average map squared value $\ev{W^2}$ are computed, we can estimate the pair counts. This can be done by numerically evaluating  the multi-dimensional integrals of Eqs.~\eqref{eq:final_RR} and \eqref{eq:DR_final} (or \ref{eq:final_DR_approx}), respectively, for $RR$ and $DR$. We tested different integration schemes, namely: 
\begin{itemize}
    \item \textsc{GSL} \citep{gough2009gnu} integration algorithm \emph{cquad;}
    \item \textsc{CUBA} library \citep{hahn2005cuba} set of optimised algorithms for multi-dimensional integration: \emph{vegas}, \emph{suave}, \emph{divonne} and \emph{cuhre.}
\end{itemize}
In each case, we need to specify a maximum tolerance on the integral relative error $\varepsilon$. The \textsc{GSL} \emph{cquad} algorithm only performs one-dimensional integrals, and we thus implement nested integrals with same $\varepsilon$ to perform the full three-dimensional integral. There are three potential sources of error associated to the analytical pair count calculation: the $p(r)$ estimation, the $\omega(\theta)$ estimation, and the precision on numerical integrals. In our methodology, the error level associated to each source is controlled respectively by the binning in $p(r)$, angular map resolution $nside$, and $\varepsilon$. In the last case however, we note that different algorithms can yield slightly different results even if $\varepsilon$ is very small.

Regarding the performances of the $RR$ implementation, for the two considered surveys and considering $6000$ bins in $(s,\mu)$, the full $RR(s,\mu)$ based on Eq.~\eqref{eq:final_RR} takes about 5-20 minutes on a single CPU using the \textsc{CUBA} library, even for $\varepsilon \approx 10^{-6}$ (\textsc{GSL} takes significantly more time when $\varepsilon < 10^{-3}$). In principle, we gain an additional factor of two when using the pair line-of-sight  mid-point definition, as in that case, there is a symmetry along the line of sight for auto-correlation and we only need to compute $\mu > 0$ pairs. The run time depends on the number of bins in $(s,\mu)$ but also in principle, on the shape of the integrand, as for complex $p(r)$ or $\omega(\theta),$ the integrals will take more time to converge (although, in our case, we did not see any noticeable difference). For the $RR$ multipole moments in Eq.~\eqref{eq:final_RR_multipoles}, the calculation only takes about $5$ seconds with \textsc{CUBA} algorithms using 30 bins in $s$, independently of the adopted value of $\varepsilon$. 

Regarding $DR$, the run times for the approximation in Eq.~\eqref{eq:final_DR_approx} are similar to those for $RR$ by definition. However, the evaluation of Eq.~\eqref{eq:DR_final} leads to large computational times of up to several weeks on a single CPU for large datasets. The run times scale linearly with the number of objects in the data sample. In that case, direct data-random pair counting might be more efficient.

The C code that follows this implementation is publicly available at \url{http://github.com/mianbreton/RR_code}. It can be used with any input $p(r)$ and $\omega(\theta)$ to predict $RR$ or $DR$ pair counts.

\subsection{Survey selection functions}

We consider two realistic redshift survey selection functions, those of SDSS-BOSS DR12 CMASS \citep{alam15} and VIPERS PDR2 \citep{scodeggio18} galaxy samples. We use the public galaxy catalogues\footnote{Available at \url{http://data.sdss.org/sas/dr12/boss/lss/} (BOSS) and \url{http://vipers.inaf.it/rel-pdr2.html} (VIPERS).} and associated angular masks. Those two samples have complementary properties and thus allow testing the method in different conditions. Indeed, while BOSS survey is wide and has a low galaxy number density, VIPERS is much narrower and denser. Each survey is composed of two separated fields on the sky but here we only consider the BOSS North Galactic Cap (NGC) and VIPERS W1 fields and we focus on $0.5<z<0.75$ and $0.7<z<1.2$ redshift intervals for BOSS and VIPERS, respectively.
\begin{figure}
\includegraphics[width=\columnwidth]{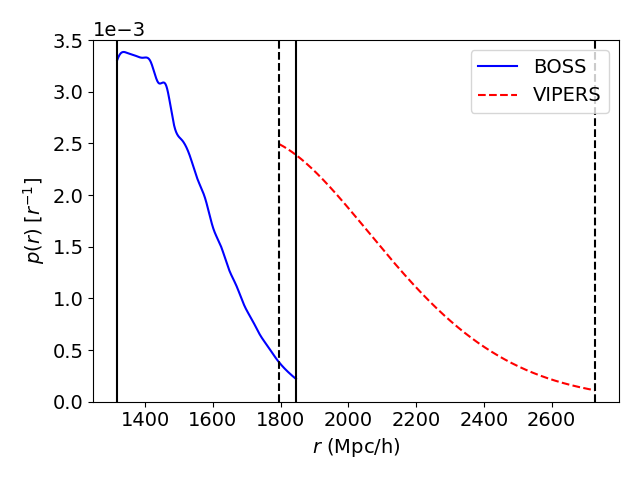} 
    \caption{Adopted radial distance distribution $p(r)$ for the BOSS NGC CMASS sample at $0.5<z<0.75$ (blue solid curve) and VIPERS W1 sample at $0.7<z<1.2$ (red dashed line). The distributions are normalised so that the integral is unity. The vertical solid (dashed) lines show the adopted sample limits for BOSS (VIPERS).}
    \label{fig:nr_surveys}
\end{figure}

The radial selection functions are shown in Fig.~\ref{fig:nr_surveys}. The BOSS $p(r)$ is estimated from the data by taking the histogram of galaxy comoving distances and cubic-spline interpolating between the bins. In the case of VIPERS, we use the fitting function for the redshift distribution given in \cite{delatorre2013vipers}. 
We assumed two cosmologies to convert redshift to comoving distance: flat $\rm \Lambda CDM$ with $\Omega_{\rm m} = 0.31$ and $\Omega_{\rm m} = 0.25$ for BOSS and VIPERS, respectively. Nonetheless, the choice of fiducial cosmology has no impact on the accuracy of the analytical predictions.   

\begin{figure}
\includegraphics[width=\columnwidth]{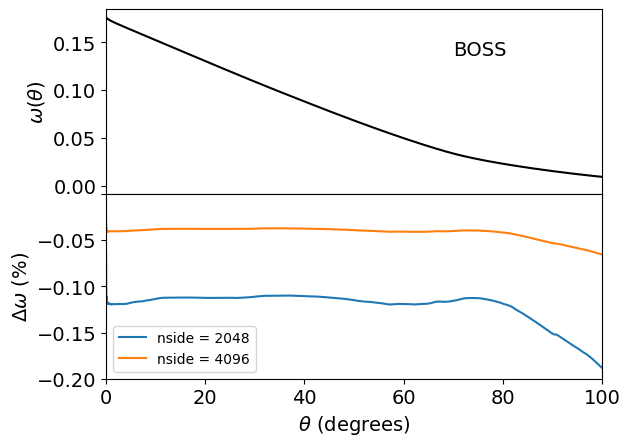}
    \caption{\emph{Top panel}: Two-point correlation function of the BOSS angular selection function obtained from a \textsc{Healpix} map with $nside = 8192$. \emph{Bottom panel}: Relative difference on the angular two-point correlation function with respect to lower resolution maps, i.e. $nside = 2048, 4096$ in blue and orange curves, respectively.}
    \label{fig:BOSS_wtheta}
\end{figure}

\begin{figure}
    \includegraphics[width=\columnwidth]{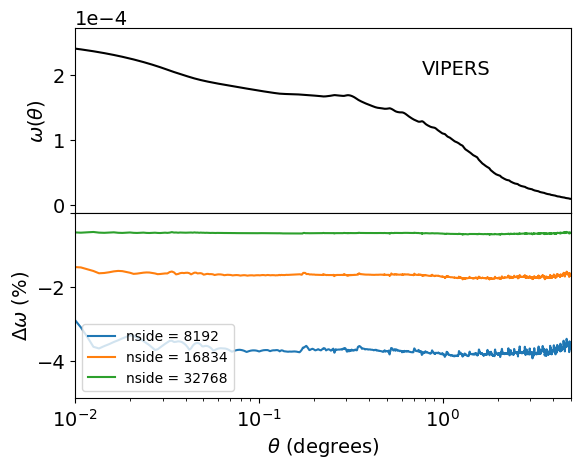} 
    \caption{\emph{Top panel}: Two-point correlation function of the VIPERS angular selection function obtained from a \textsc{Healpix} map with $nside = 65536$. \emph{Bottom panel}: Relative difference on the angular two-point correlation function with respect to lower resolution maps, i.e. $nside = 8192, 16834, 32768$ in blue, orange, and green curves, respectively.}
    \label{fig:VIPERS_wtheta}
\end{figure}

\begin{figure}
\centering
\includegraphics[width=\columnwidth]{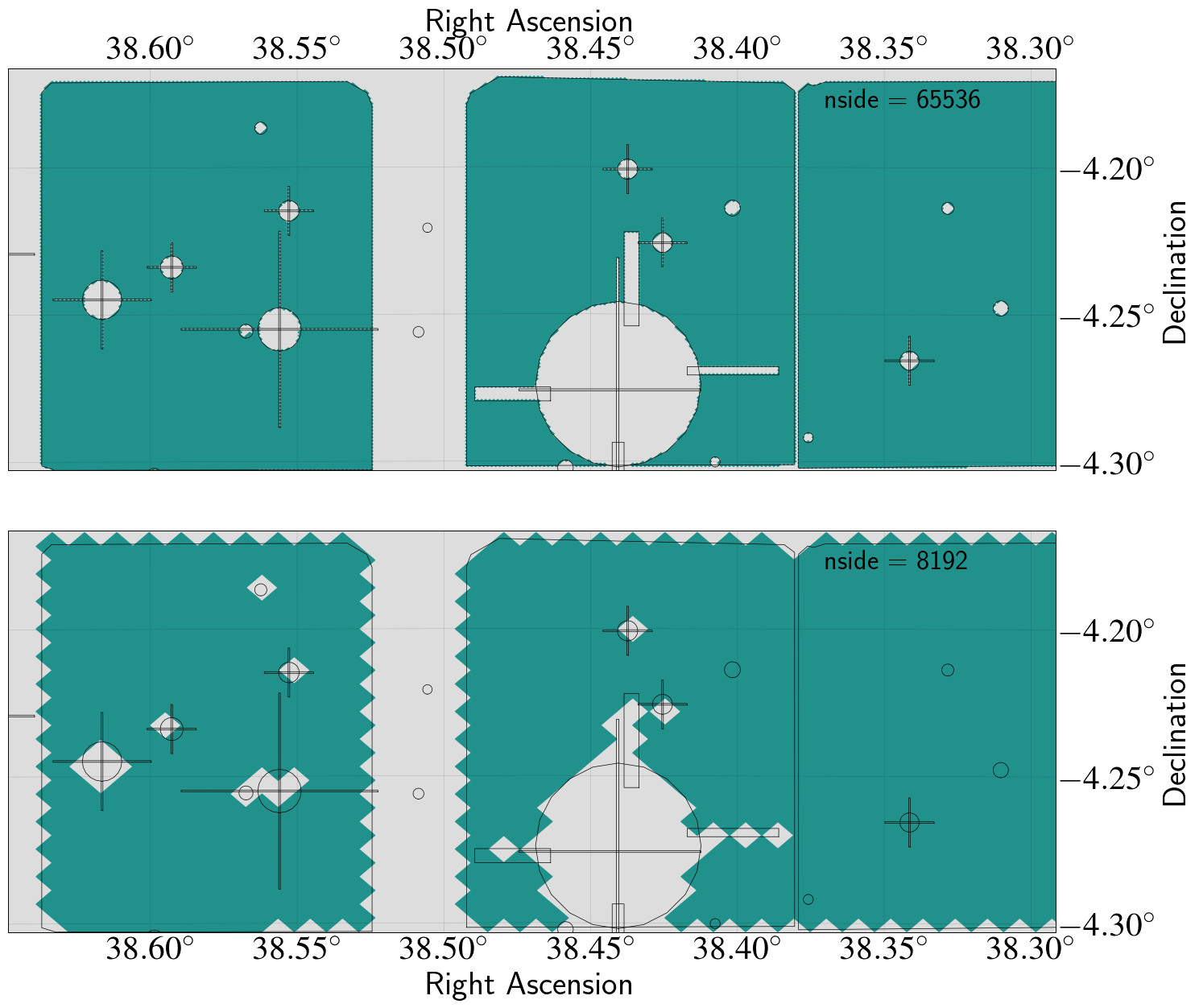} 
    \caption{Detail of the VIPERS W1 angular mask showing the impact of \textsc{Healpix} resolution on the sampling of the survey angular mask.}
    \label{fig:vipers_resolution}
\end{figure}

The angular selection functions that we used for VIPERS W1 and BOSS NGC are presented in Appendix~\ref{sec:appendix_footprints}. In total, the surface covered by the BOSS NGC (VIPERS W1) field is 7427.4~deg$^2$ (10.7~deg$^2$). The angular selection function enters in Eq.~\eqref{eq:final_RR} through its auto-correlation function. The latter are given in Figs.~\ref{fig:BOSS_wtheta} and~\ref{fig:VIPERS_wtheta}, respectively for BOSS and VIPERS.
We test different map resolutions by varying the {\sc Healpix} resolution parameter $nside$ from $2048$ to $8192$.
For BOSS, the correlation function is very smooth. The relative difference between
$nside = 2048, 4096$ cases and $nside = 8192$ is roughly constant, at $0.1\%$ and $0.05\%$ respectively. In the case of VIPERS, the angular mask has more small-scale features but similarly, the relative differences between angular correlation functions based on different map resolutions are nearly constant in scale. The bias is larger than in the BOSS case, with a relative difference with respect to $nside = 65536$ of $4\%$, $2\%,$ and $0.5\%$, respectively for $nside = 8192, 16834, 32768$. This difference is due to the fact that we need a significantly higher map resolution to correctly account for the angular selection function as illustrated in Fig.~\ref{fig:vipers_resolution}. The latter figure shows a detail of the VIPERS angular mask pixelated at $nside=8192$ and $nside=65536$. Overall, the convergence of the angular correlation function with increasing $nside$, allows us to assess that resolutions of $nside = 8192$ and $nside = 65536$ are sufficiently accurate for BOSS and VIPERS respectively.
We note that while the map resolution impacts the estimation of the angular selection function two-point correlation function, it also changes the estimation of $n(r)$ through $\ev{W}$, which partly compensates the bias from $\omega(\theta)$ in the final $RR$ and $DR$ estimations.

\subsection{$RR$ counts}
\label{sec:results_RR}
\begin{figure}
\includegraphics[width=0.96\columnwidth]{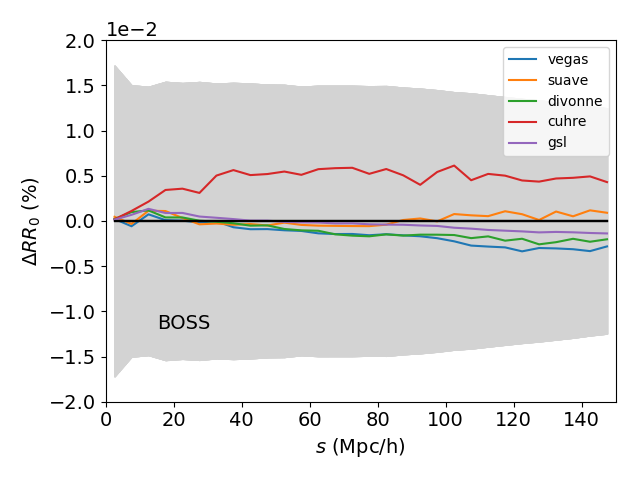} 
\includegraphics[width=0.96\columnwidth]{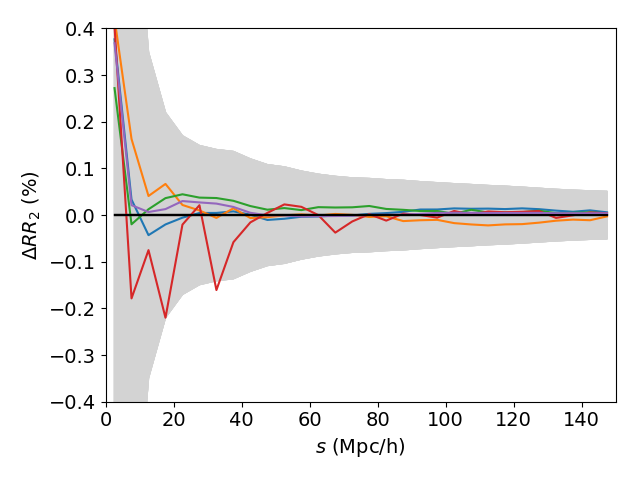} 
\includegraphics[width=0.96\columnwidth]{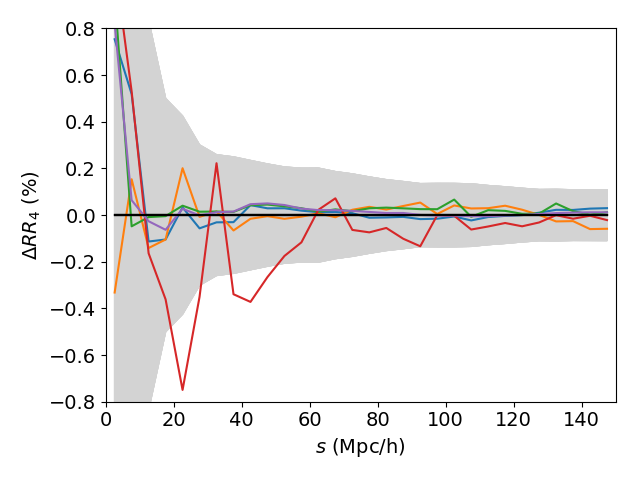} 
    \caption{Relative difference between the analytical and random catalogue-based mean $\left<RR\right>$  pair-count multipole moments ($\ell = 0,2,4$, in the top, middle, and bottom panels, respectively) for BOSS. The grey shaded area shows the standard deviation among the random catalogues, while blue, orange, green, red, and purple curves present the relative differences obtained with \emph{vegas}, \emph{suave}, \emph{divonne}, \emph{cuhre,} and \emph{gsl} algorithms, respectively, when using $\varepsilon = 10^{-5}$.}
    \label{fig:BOSS_RR}
\end{figure}

\begin{figure}
\includegraphics[width=0.96\columnwidth]{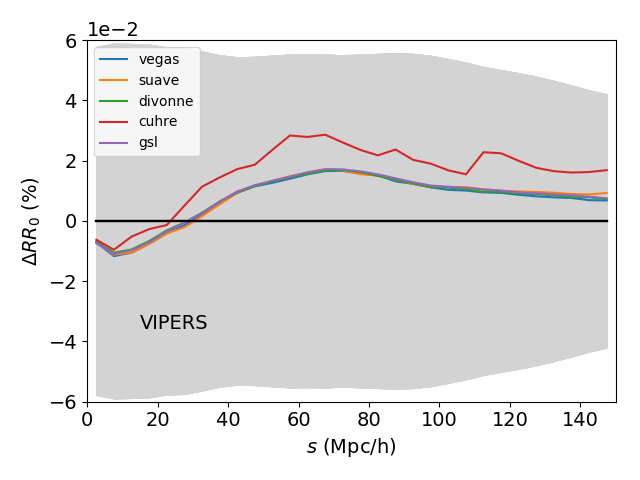} 
\includegraphics[width=0.96\columnwidth]{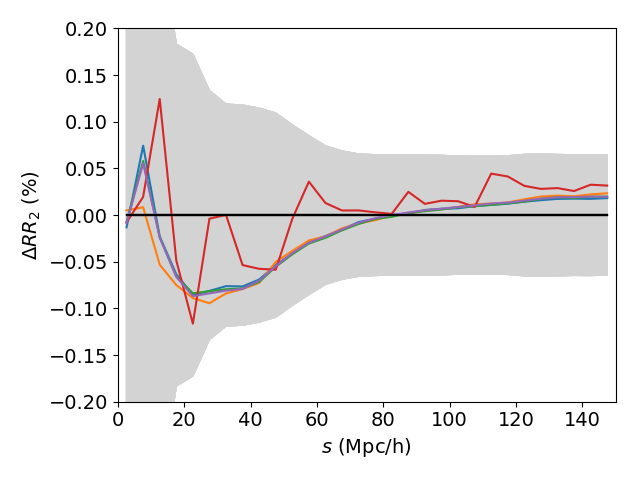} 
\includegraphics[width=0.96\columnwidth]{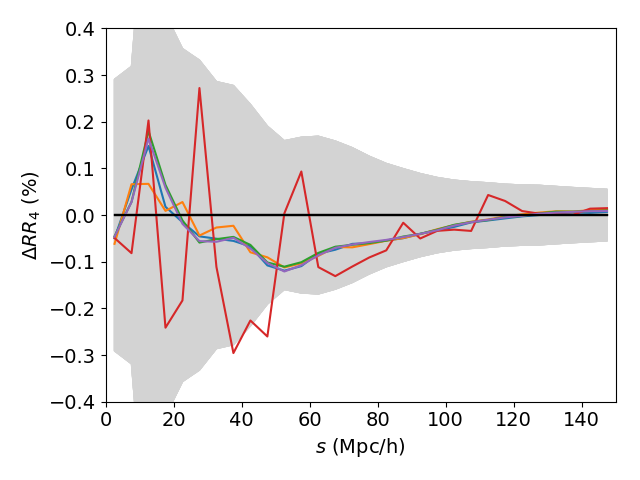} 
    \caption{Same as Fig.~\ref{fig:BOSS_RR}, but for VIPERS.}
    \label{fig:VIPERS_RR}
\end{figure}

We compare our analytical prediction for $RR$ with the average random-random counts $\ev{RR}$, obtained from $100$ random samples constructed using the same radial and angular selection functions. Within the considered redshift intervals, there are $435185$ BOSS and $24316$ VIPERS galaxies and we generate $3\times10^{7}$ and $3.9\times10^{6}$ points per random sample, respectively (i.e. multiplicative factors of about 70 and 160 with respect to the data) with the radial distributions shown in Fig.~\ref{fig:nr_surveys}. We compute the pair counts from the random samples using the fast \textsc{Corrfunc} pair-counting code \citep{sinha2020corrfunc}. Our method predicts anisotropic $RR(s,\mu)$ counts, but to simplify the comparison, we consider the first three even multipoles, that is, $RR_\ell(s) = (2\ell + 1)/2 \sum_{i} RR(s,\mu_i) \mathcal{L}_\ell(\mu_i)\Delta \mu$ with $\ell = 0,2,4$,  where linear bins $\mu_i$ extend from $-1$ to $1$. Those comparisons are presented in Fig.~\ref{fig:BOSS_RR} for BOSS and in Fig.~\ref{fig:VIPERS_RR} for VIPERS.
We see that for both surveys, the relative difference between the analytical computation and $\left<RR\right>$ is well within the variance of the random samples. We compare the results obtained with different numerical integration algorithms (see figure insets and Section \ref{sec:numerical_implementation}) and find that \emph{cuhre} tends to depart from the others algorithms, which is understandable since it is intrinsically different from the others. 
If we ignore \emph{cuhre}, we see that, at most, the relative difference between the analytical computation and $\left<RR\right>$ remains within $3\times10^{-5}$ for BOSS and $1.7\times10^{-4}$ for VIPERS on the monopole, and about $10^{-3}$ for the quadrupole and hexadecapole for both surveys.

The variance on the random sample counts depends on the number of points in the sample and, thus, we may ask what is the number of random points needed to achieve the same accuracy as in the analytical method. \cite{keihanen2019estimating} showed that the relative variance on $RR$ in a given bin is:
\begin{equation}
    \textrm{var}(RR) = \frac{2}{N_r(N_r-1)}\left\{ 2(N_r-2)\left[ \frac{G^t}{(G^p)^2} - 1\right] + \frac{1}{G^p}- 1 \right\}
,\end{equation}
with $N_r$ the number of random points, and $G^p$, $G^t$ terms are \citep{landy1993bias}:
\begin{align}
    G^p =& \frac{\ev{n_p}}{N_r(N_r-1)/2},\\
    G^t =& \frac{\ev{n_t}}{N_r(N_r-1)(N_r-2)/2},
\end{align}
with $\ev{n_p}$ and $\ev{n_t}$ the number of pairs and triplets averaged over several realisations. While $G^p$ can easily be estimated from the random samples, we directly solve for $G^t$ from the estimated var($RR$). We can then deduce which $N_r$ give standard deviations similar to $3\times10^{-5}$ and $1.7\times10^{-4}$ for the monopole. We found that we need an additional factor of at least 20 (10) for BOSS (VIPERS) in the number of random points. Therefore, the analytical method allows the achievement of the same accuracy as by using a random sample with about $20\times70$ ($10\times160$) more points than data in BOSS (VIPERS). Finally, we note that \textsc{CUBA} integration algorithms have parameters that can be potentially further fine-tuned to achieve better accuracy.

\subsection{$DR$ counts}

\begin{figure}
\includegraphics[width=\columnwidth]{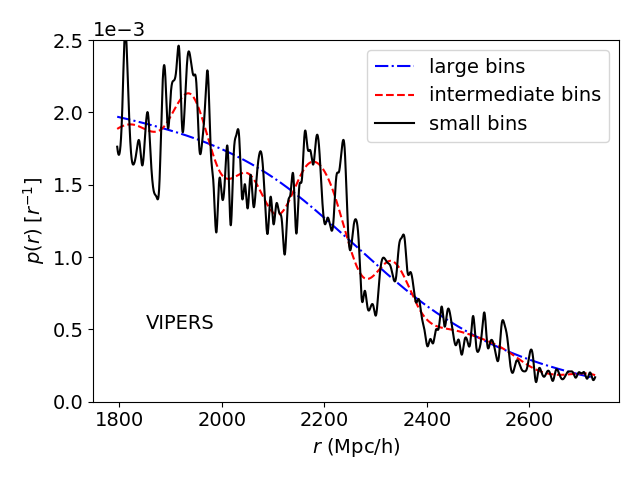} 
\caption{Estimated radial distance distribution $p(r)$ in the VIPERS sample at $0.7<z<1.2$ using different linear bin size in $r$. The distributions are normalised so that the integral is unity. The blue dotted-dashed, red dashed, and black solid lines are the distributions obtained when using large, intermediate, and small bin sizes, respectively.}
    \label{fig:nr_VIPERS_bins}
\end{figure}

\begin{figure}
\includegraphics[width=\columnwidth]{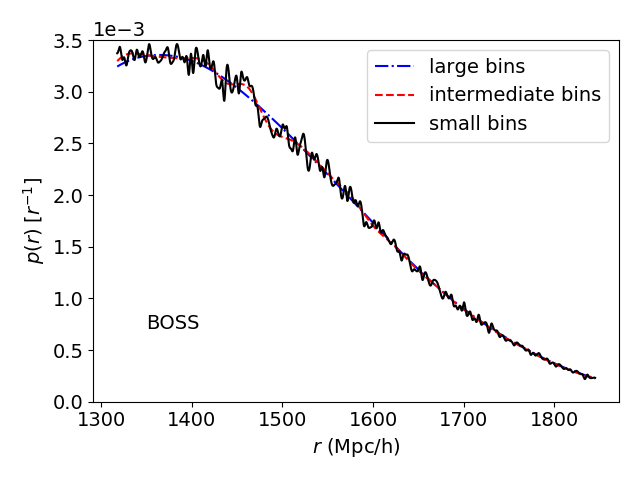} 
    \caption{Same as Fig. \ref{fig:nr_VIPERS_bins}, but for BOSS.}
    \label{fig:nr_BOSS_bins}
\end{figure}

In the $DR$ case, we need to rely on approximations.
Under the approximation in Eq.~\eqref{eq:approx_DR}, we have two possibilities to calculate $DR$ counts: either a discrete sum over all source distances as in Eq.~\eqref{eq:DR_final} or by further approximating the discrete sum by an integral as in Eq.~\eqref{eq:final_DR_approx}. In the last case, we can already anticipate that the results will depend on the input $p(r_1)$, particularly its ability to reproduce line-of-sight structures in the data. In Figs.~\ref{fig:nr_VIPERS_bins} and \ref{fig:nr_BOSS_bins}, we show different estimations of the data $p(r_1)$ in VIPERS and BOSS, varying the bin size in $r_1$. In the limit where $p(r_1)$ resembles a sum of Dirac delta functions, Eq.~\eqref{eq:final_DR_approx} should be equivalent to Eq.~\eqref{eq:DR_final}. For the random part we use in $p(r_2)$ the distributions provided in Fig.~\ref{fig:nr_surveys}. It is worth noting that we use cubic splines to model the data distributions in Figs.~\ref{fig:nr_VIPERS_bins} and \ref{fig:nr_BOSS_bins}. While other ways of estimating $p(r_1)$ could have been chosen, we only focus on the relative importance of the binning, and therefore, the method used for the estimation is irrelevant here (however, it would be necessary for an accurate, in-depth characterisation of the radial selection function).

\begin{figure}
\includegraphics[width=0.96\columnwidth]{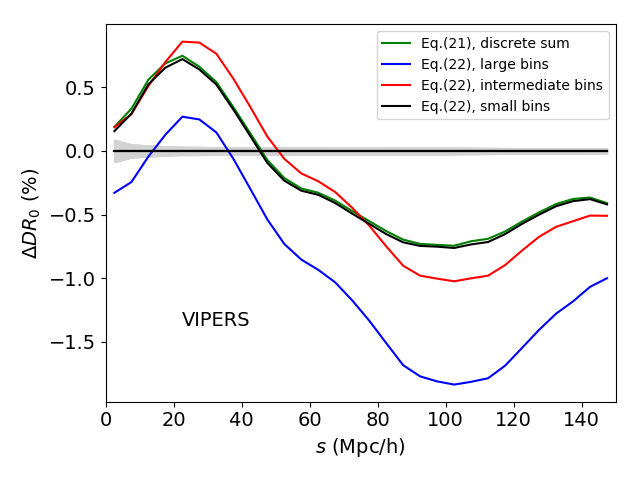} 
\includegraphics[width=0.96\columnwidth]{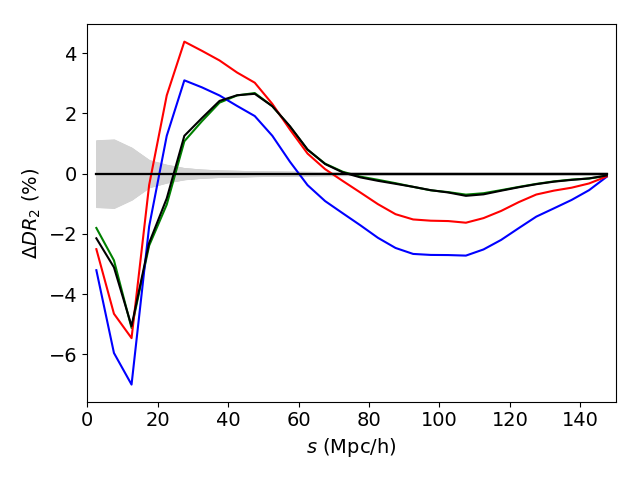} 
\includegraphics[width=0.96\columnwidth]{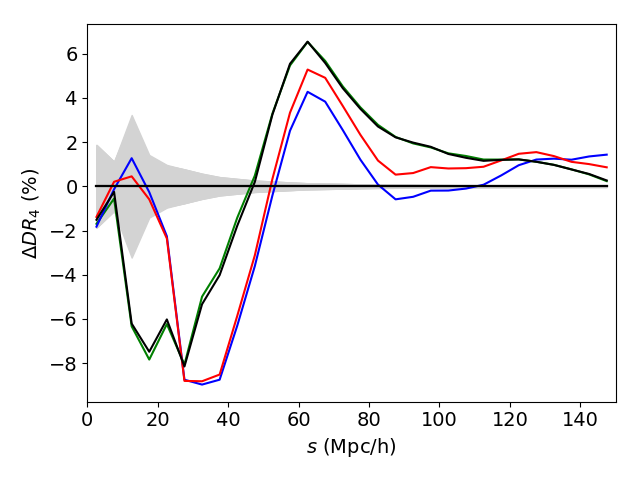} 
    \caption{Relative difference between the analytical and random catalogue-based mean $\left<DR\right>$  multipoles ($\ell = 0,2,4$, in the top, middle, and bottom panels, respectively) for VIPERS. The grey shaded area shows the standard deviation among the random catalogues. The predictions use Eq.~\eqref{eq:approx_DR} and $p(r)$ with large (blue), intermediate (red), and small (black) bin sizes. These use \emph{vegas} with $\varepsilon = 10^{-5}$. The green line shows the prediction of Eq.~\eqref{eq:DR_final} obtained with GSL using $\varepsilon = 10^{-4}$.}
    \label{fig:VIPERS_DR}
\end{figure}

Following the same methodology as in Section~\ref{sec:results_RR}, we compute $\ev{DR}$ for both surveys using the same data catalogue and 100 random samples, which we later compare to the predictions based on Eqs.~\eqref{eq:DR_final} and \eqref{eq:final_DR_approx} using different input data $p(r_1)$.
In the case of VIPERS, we find that when using Eq.~\eqref{eq:DR_final}, the discrepancy between the analytical prediction and direct pair counting is on the order of 1\% for the monopole, up to several percent for the quadrupole and hexadecapole, as shown in Fig.~\ref{fig:VIPERS_DR}. Moreover, we see that the prescription in Eq.~\eqref{eq:final_DR_approx} leads to a systematic bias of up to about $2\%$ on the monopole when using a large binning in the input $p(r_1)$, but converges towards Eq.~\eqref{eq:DR_final} result when a small binning is adopted, as expected.

\begin{figure}
\includegraphics[width=0.96\columnwidth]{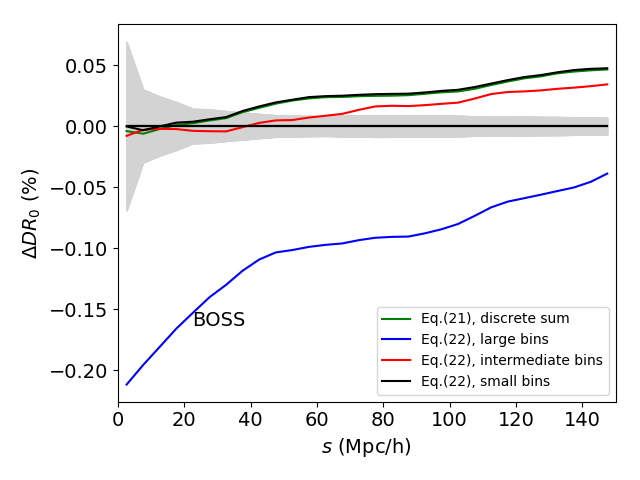} 
\includegraphics[width=0.96\columnwidth]{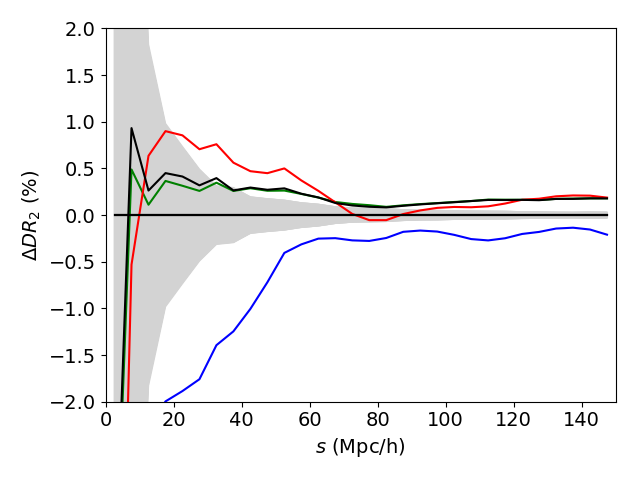} 
\includegraphics[width=0.96\columnwidth]{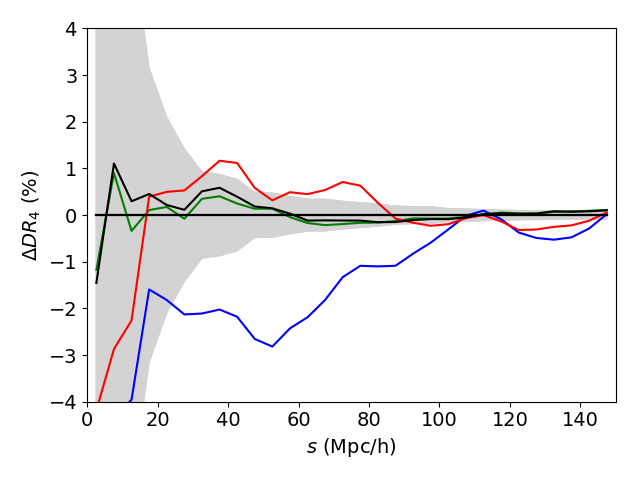} 
    \caption{Same as Fig~\ref{fig:VIPERS_DR} but for BOSS. Here. the prediction of Eq.~\eqref{eq:DR_final} in green is obtained with GSL using $\varepsilon = 10^{-3}$.}
    \label{fig:BOSS_DR}
\end{figure}

In the case of BOSS, we find similar trends but with an higher accuracy, as shown in Fig.~\ref{fig:BOSS_DR}. We find at most a difference of $5\times10^{-4}$ on the monopole between the analytical solution, either Eq.~(\ref{eq:DR_final}) or Eq.~(\ref{eq:final_DR_approx}) with a fine $p(r_1)$, and direct pair counting. Regarding the quadrupole and hexadecapole, the relative difference is about $1\%$. Here, the approximation in Eq.~\eqref{eq:approx_DR} is more appropriate since the data sample is larger. This explains the improved accuracy that is reached.
We emphasise that the variance in Figs.~\ref{fig:VIPERS_DR} and~\ref{fig:BOSS_DR} only comes from the random samples since a single data catalogue is used. Therefore, increasing the number of random points would reduce this variance. 
Overall, because of the approximation in Eq.~\eqref{eq:approx_DR}, our analytical $DR$ predictions remain biased, exceeding the typical variance introduced by random sampling in direct pair counting.

\section{Conclusion}
\label{sec:conclusion}

In this paper, we present general analytical expressions for the random-random and data-random pair counts in the case of a realistic survey geometry. The main results are given in Eq.~\eqref{eq:final_RR} (or Eq.~\ref{eq:final_RR_multipoles} for the multipole moments) for $RR$ and in Eq.~\eqref{eq:final_DR_approx} for $DR$. These expressions can be solved numerically in an efficient way. This method, which does not rely on generating random mocks, only takes as input the comoving radial distance distribution in an assumed cosmology and the angular selection function two-point correlation function, which only needs to be estimated a single time for a given survey. Once those quantities are provided, the full computation takes about a few minutes to obtain anisotropic pair counts $RR(s,\mu)$ and a few seconds for its multipole moments, using a single CPU and standard libraries for three-dimensional integration. 

We tested this method in the context of the BOSS and VIPERS survey geometries and found excellent agreements with expected $RR$ pair counts. The predicted counts exhibit a high accuracy for the cases investigated in this work, equivalent to that we would obtain by performing pair counting in random samples of about 1400-1600 more random points than data in those surveys for the monopole. The main advantage is that the method is fast and does not rely on any spatial sampling, while usually we need to generate a random catalogue with at least 50 times the number of objects in the data. We believe that this can be of some use for future surveys with large data samples and very expensive $RR$ pair count calculations.

The $DR$ pair counts can also be calculated analytically based on certain approximations. We found that the results are slightly biased with respect to the expected counts. For VIPERS and BOSS, we found a bias with respect to direct pair counts of 1\% and 0.05\%, respectively, for the monopole, up to several percents on the quadrupole and hexadecapole. This bias should decrease with the increasing number of data points. When estimating $DR$ for several data samples, we need to compute, for each sample, its angular two-point correlation function with respect to the survey angular selection function.

Overall, the method presented in this paper for efficiently evaluating the survey window two-point function should be very useful when dealing with massive galaxy surveys. The  formulae provided are fast in terms of the speed of the evaluation. With further efficient parallelisation \citep[e.g.][]{hahn2015concurrent}, we should be able to compute $RR$ and $DR$ in an extremely small amount of time. In that case, we could imagine $RR$ and $DR$ being evaluated in different cosmologies at each step of a cosmological likelihood analysis. This opens up new horizons for the way we analyse galaxy survey data in the future.


\begin{acknowledgements}
We thank Eric Jullo for his help on dealing with partial, high-resolution {\sc Healpix} maps and his comments on the draft. 

We thank the Instituto de Astrofísica de Andalucía (IAA-CSIC), and the Spanish academic and research network (RedIRIS, \url{http://www.rediris.es}) in Spain for providing the skun@IAA\_RedIRIS server that allowed us to run the calculations for high-resolution $nside=65535$ maps.

This work has been carried out thanks to the support of the OCEVU Labex (ANR-11-LABX-0060) and of the Excellence Initiative of Aix-Marseille University - A*MIDEX, part of the French “Investissements d’Avenir” programme. 

Funding for SDSS-III has been provided by the Alfred P. Sloan Foundation, the Participating Institutions, the National Science Foundation, and the U.S. Department of Energy Office of Science. The SDSS-III web site is http://www.sdss3.org/.

SDSS-III is managed by the Astrophysical Research Consortium for the Participating Institutions of the SDSS-III Collaboration including the University of Arizona, the Brazilian Participation Group, Brookhaven National Laboratory, Carnegie Mellon University, University of Florida, the French Participation Group, the German Participation Group, Harvard University, the Instituto de Astrofisica de Canarias, the Michigan State/Notre Dame/JINA Participation Group, Johns Hopkins University, Lawrence Berkeley National Laboratory, Max Planck Institute for Astrophysics, Max Planck Institute for Extraterrestrial Physics, New Mexico State University, New York University, Ohio State University, Pennsylvania State University, University of Portsmouth, Princeton University, the Spanish Participation Group, University of Tokyo, University of Utah, Vanderbilt University, University of Virginia, University of Washington, and Yale University.

\end{acknowledgements}

\bibliographystyle{aa}
\bibliography{biblio} 

\begin{appendix}
\onecolumn

\section{VIPERS and SDSS-BOSS survey footprints}
\label{sec:appendix_footprints}
\begin{figure*}[h]
\centering
\includegraphics[width=\columnwidth]{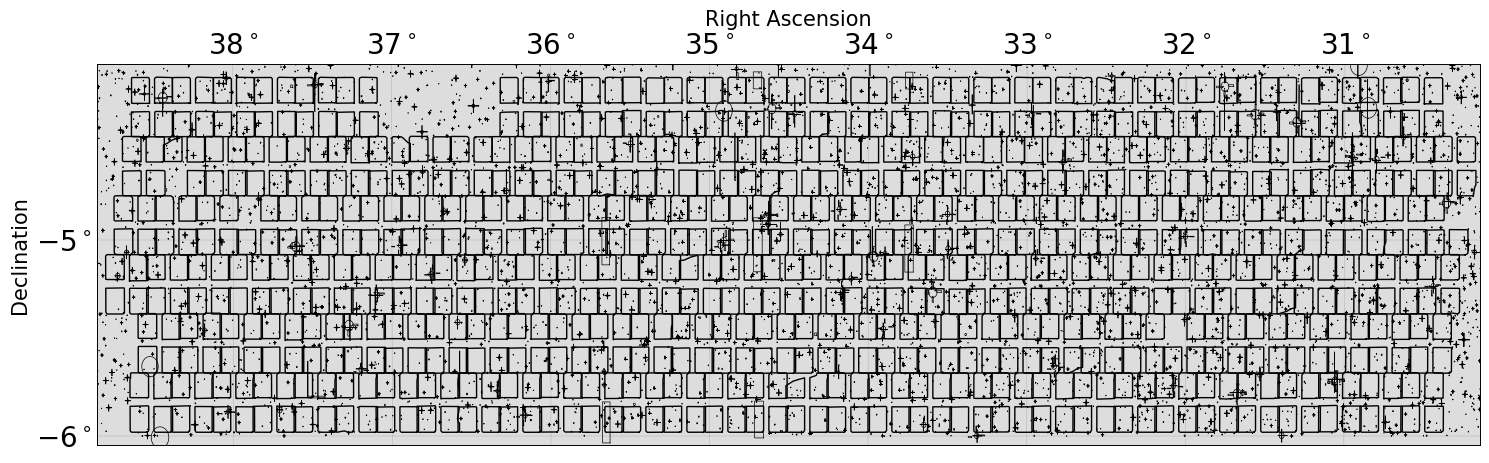} 
    \caption{VIPERS W1 footprint.}
    \label{fig:vipers_footprint}
\end{figure*}

\begin{figure*}
\centering
\includegraphics[width=\columnwidth]{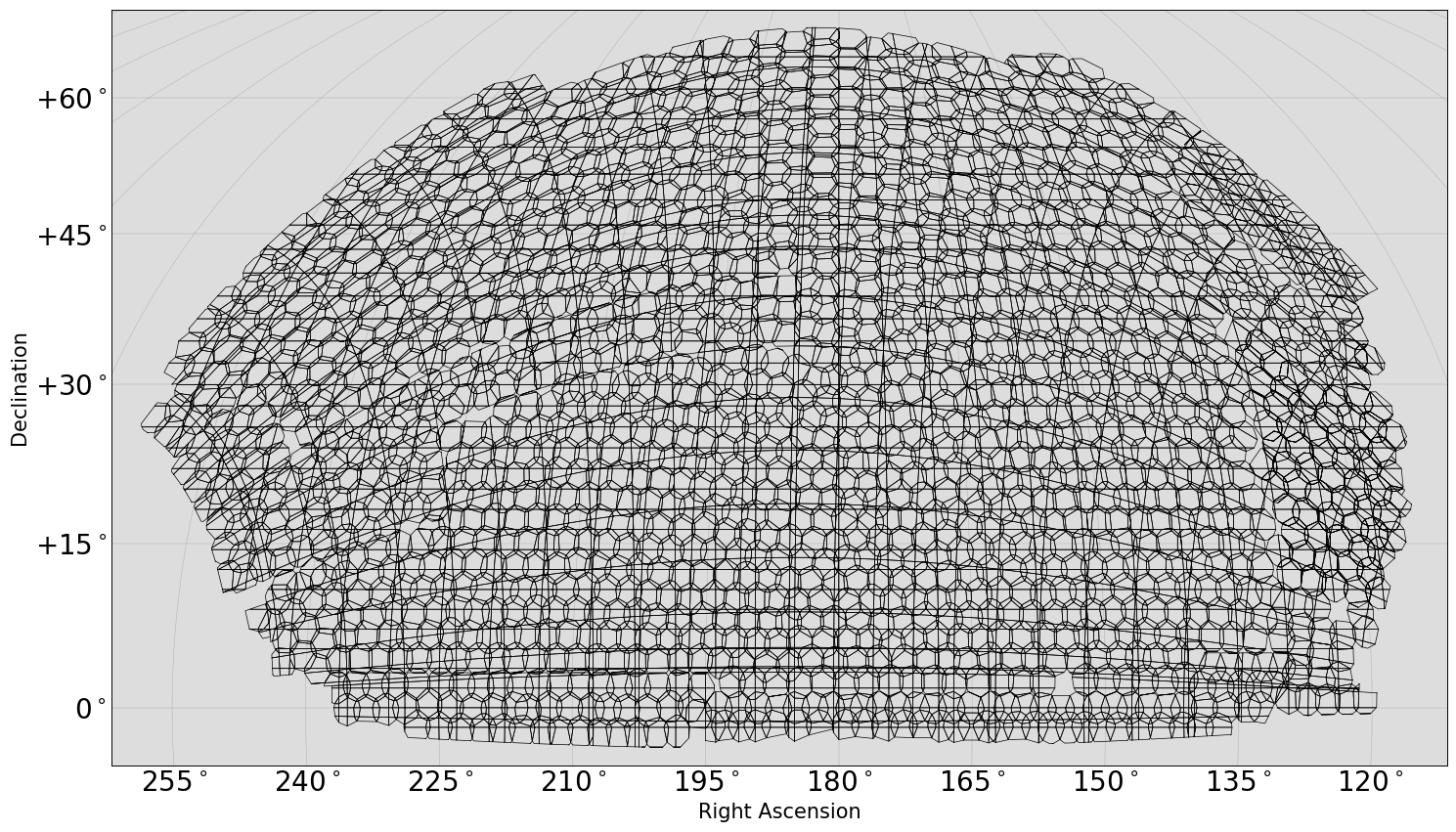} 
    \caption{BOSS CMASS NGC footprint.}
    \label{fig:boss_footprint}
\end{figure*}

In Figs.~\ref{fig:vipers_footprint} and \ref{fig:boss_footprint}, we provide the footprints and angular masks for VIPERS W1 and SDSS-BOSS CMASS NGC fields, respectively, which we used in this analysis. In the case of BOSS angular mask, each distinct mask polygon has an associated tiling success rate, which is a measure of the completeness in associating fibres to potential spectroscopic targets in the survey. We use this quantity as a weight in defining the angular selection function. In the case of VIPERS, the angular selection function is taken to be unity inside the spectroscopic mask (quadrant-shaped polygons) and null otherwise, except in the regions of the photometric mask (circular- and star-shaped polygons), where it is also set to zero.


\end{appendix}
\end{document}